\documentclass[twocolumn,english,aps,prd, floatfix,amssymb,superscriptaddress,a4paper]{revtex4-1}
\usepackage{lipsum}
\usepackage{amsmath}

\usepackage{amssymb}
\usepackage{graphicx}
\usepackage{physics}
\usepackage{slashed}
\usepackage{hyperref}
\usepackage[dvipsnames]{xcolor}
\usepackage{float}
\usepackage[caption=false]{subfig}

\begin{document}

\title{{Tunable tachyon mass in the $PT$-broken massive Thirring model} }  

\author{Benjamin Liégeois}
\affiliation
{Institute for Theoretical Physics, ETH Zurich, 8093 Zurich, Switzerland}

\author{Chitra Ramasubramanian}
\affiliation
{Institute for Theoretical Physics, ETH Zurich, 8093 Zurich, Switzerland}

\author{Nicolò Defenu}
\affiliation
{Institute for Theoretical Physics, ETH Zurich, 8093 Zurich, Switzerland}


\date{\today}

 \begin{abstract}
We  study the full phase diagram of  a non-Hermitian  PT-symmetric generalization of the  paradigmatic two-dimensional massive Thirring model.  Employing the nonperturbative functional renormalization group, we find that the model hosts a regime where PT symmetry is spontaneously broken. This new phase  is characterized by  a relevant imaginary mass, corresponding to monstronic excitations displaying exponentially growing amplitudes for timelike intervals and tachyonic (Lieb-Robison-bound breaking, oscillatory) excitations for spacelike intervals. Furthermore, since the phase manifests itself as an unconventional attractive spinodal fixed point, which is typically unreachable in finite real-life systems, we find that the effective renormalized mass reached can be tuned through the microscopic parameters of the model. Our results further predict that the new phase is robust to  external gauge fields, contrary to the celebrated BKT phase in the PT unbroken sector. The  gauge field  then provides an effective and easy means to tune the renormalized imaginary mass through a wide range of values, and therefore the amplitude growth/oscillation rate of the corresponding excitations. 
\end{abstract}

\maketitle

\section{Introduction} 

The study of quantum-mechanical systems has largely relied on the assumption of Hermiticity of Hamiltonians. This effectively ensures the reality of their spectrum and guaranties the conservation of probability. Real systems nevertheless tend to interact with their environment leading to nonunitary evolutions which can be described in terms of non-Hermitian effective Hamitonians.  It was however realized that non-Hermiticity does not necessarily invalidate the essential properties of Hermitian Hamitonians. In particular, a broad class of non-Hermitian matrices termed pseudo-Hermitian was shown to exhibit real spectra\,\cite{Mostafazadeh_1,Mostafazadeh_2,Mostafazadeh_3}. While this typically includes Hamitonians possessing a range of antilinear symmetries\,\cite{Bender2002}, one of them stands out for its experimental accessibility: PT symmetry\,\cite{Bender_PT_Real}, i.e. simultaneous reversal of space and time. A Hamiltonian $H$ is PT-symmetric if it commutes with the product operator $PT$ where $P$ and $T$ are the parity and time-reversal operators, while not necessarily commuting with $P$ or $T$ individually.

An intuitive way to picture a PT-symmetric system is to consider two subsystems related through spatial inversion handled in such a way that the gains encountered by the former correspond exactly to the losses experienced by the latter\,\cite{BenderBook}. In this picture, one naturally understands how such a symmetry can be implemented through spatial engineering of gain-loss structures. In addition to its experimental accessibility, this class of non-Hermitian Hamitonians often displays a rich distinctive feature, namely the possibility of breaking PT symmetry spontaneously\,\cite{NH_Ashida,BenderBook}. PT symmetry is said to be unbroken if every eigenstate of the Hamitonian $H$ is an eigenstate of the PT operator. In this case, the spectrum of $H$ is entirely real despite its non-Hermiticity. In contrast, PT symmetry is spontaneously broken if some eigenstates have complex eigenvalues. These then occur in complex conjugate pairs. The spontaneous breaking of PT symmetry is generally associated with the coalescence of eigenvalues and their respective eigenstates at an exceptional point \cite{Kato1995} in the discrete spectrum or at spectral singularities\,\cite{SpectralSingularity} in the continuum. Altogether, this makes PT symmetry an extraordinary platform to explore novel critical behaviors in non-Hermitian systems. 

 While in the past research in non-Hermitian systems has largely focused on single and effective few-body problems, recent technological advances  in the design of open many-body systems in ultracold atoms and exciton-polariton condensates~\cite{RealisePTQMBS, Exp_2, Exp_3, Exp_5, Exp_6, Ref_PseudoYL_Exp1, Ref_PseudoYL_Exp2,Ref_PseudoYL_Exp3,Ref_PseudoYL_Exp4}  have opened  a new arena for strongly interacting non-Hermitian systems and in particular non-Hermitian quantum critical phenomena. Historically, the study of phase transitions in non-Hermitian systems started out with the investigation of the Lee-Yang edge singularity, where the Ising model in the presence of an imaginary magnetic field was demonstrated to present a universal scaling, connected to its equilibrium criticality\,\cite{LeeYang,Yang-Lee_Phi3, FRG_Phi3_An}. While these predictions were long viewed as purely of theoretical interest, the aforementioned recent experimental developments have opened the door to the experimental exploration of a wide variety of collective non-Hermitian phenomena\,\cite{Ref_PseudoYL_1,Ref_PseudoYL_2,Ref_PseudoYL_3,Ref_PseudoYL_4, Ref_PseudoYL_5, Ref_PseudoYL_6}. Specifically, Yang-Lee edge singularities have been detected in recent experiments on a normal metal connected to superconducting leads\,\cite{Exp_LY_4,Exp_LY_2,Exp_LY_3}. 
 
Recent studies have predicted a host of phenomena in non-Hermitian many-body systems, including a many-body localization-delocalization transition\,\cite{Ref_PseudoYL_7},  dynamical phase transitions\,\cite{syed2022universal}, continuous quantum phase transitions unaccompanied by gap closures\,\cite{NoGapClosing_LiebRobinson} and Kibble-Zurek scaling across exceptional points\,\cite{Ref_PseudoYL_9}.  
  Interplays between the  spontaneous breaking of both PT and  a continuous symmetry have also been explored \cite{NH_SSB_Goldstone,Conservation_Ellis,NH_SSB_Golstone2, NH_SSB_Goldstone_Mannheim, Superfluid_2_Ellis}. PT symmetry breaking may also occur in concurrence with topological scaling phenomena as was predicted in a non-Hermitian generalization of the sine-Gordon model~\cite{Duality}, a model of profound relevance to both condensed matter and high-energy physics\,\cite{Ref_PseudoYL_8,Ashida_PTSG}.  Such imaginary-coupled sine-Gordon model was formally predicted to host excitations with amplitudes that grow exponentially in time, termed monstrons\,\cite{Monstron_cFunction_SG}. Perturbative RG has flagged the presence of semicircular flows breaking the $c$-theorem~\cite{Zamolodchikov_cTheorem} in the PT symmetry broken phases of the  generalized sine-Gordon model~\cite{Ashida_PTSG} and non-Hermitian Kondo models~\cite{PT_Kondo1,PT_Kondo2}. Clearly,  these 
results  warrant an in-depth revision of the standard notions of critical behavior and universality.

In this paper, we investigate the interplay between topological scaling and PT symmetry breaking in the
non-Hermitian generalization of the massive Thirring model in (1+1)-dimensions~\cite{Duality}.  

The paradigmatic original Hermitian  model is dual to the quantum sine-Gordon model~\cite{coleman1975quantum} and manifests the Berezinskii-Kosterlitz-Thouless (BKT) transition\,\cite{Russian_FRGThirring}. While the duality was predicted to extent to the corresponding non-Hermitian generalization in the PT-unbroken phase~\cite{Duality}, it may not necessarily hold in the presence of spontaneous breaking of PT-symmetry.


To study the generalization of the Thirring model, we employ the functional renormalization group (FRG), a versatile formalism based on the effective action paradigm\,\cite{Delamotte_2012, Gies_2012, Wetterich_1993, Berges_FRG}.
FRG  has  recently been used to study interactions in the non-Hermitian setting~\cite{ZambelliZanusso,FRG-NH-Vertex, Bender_FRG, AshidaThesis}. We obtain the full nonperturbative flow diagram of the model.  
Our FRG analysis predicts a new phase displaying a relevant imaginary mass, corresponding for spacelike intervals to tachyonic (Lieb-Robison bound breaking, oscillatory) excitations, or excitations displaying exponentially growing amplitudes for timelike intervals called monstrons. This novel phase manifests itself as an unconventional attractive spinodial fixed point. Consequently, the effective renormalized mass can be tuned through the microscopic parameters of the model.

It was shown that the BKT transition in the original Hermitian model was wiped out by the presence of external gauge fields~\cite{Russian_FRGThirring}. It is therefore of great interest to study their influence on the extended phase diagram of the PT-symmetric non-Hermitian generalization of the massive Thirring, and in particular on the PT-broken sector. The external gauge fields are relevant to many physical systems: they can notably be realized as incommensurabilities in classical systems~\cite{Incomm, Pert_RG_Incomm}, magnetic fields in spin systems~\cite{ChitraSpinB} soliton fugacities~\cite{Soliton_density} in bosonic settings or chemical potentials~\cite{Schutz_LatticeFermions} in fermionic systems. While our results further predict a dramatic change in the renormalization group flows in the presence of an external gauge field,  the new tachyonic phase is robust and the gauge field is shown to provide an additional practical way to tune the corresponding imaginary mass.

\vspace{-5mm}



\section{Generalized massive Thirring}

We  consider  the  generalized massive Thirring model in $(1+1)$-dimensional space-time~\cite{Duality}, whose Hamiltonian is given by 
\begin{equation}\label{genThirring}
{H}=\int d x\left[\bar{\psi}\left( -i\slashed{\nabla}+  {m}_1+{m}_2\gamma_5 \right) \psi+\lambda (\bar{\psi} \psi)^{2}\right],
\end{equation}
where $\bar{\psi}({x}, t)=\psi^{\dagger}({x}, t) \gamma_{0}$. With the following conventions $\gamma_{0}$ $=\sigma_{1}$ , $\gamma_{1}=i \sigma_{2}$ and $\gamma_{5}=\gamma_{0} \gamma_{1}=\sigma_{3}$, where $\sigma$ are the Pauli matrices,  we have $\gamma_{0}^{2}=\gamma_{5}^2=1$ and $\gamma_{1}^{2}=-1$.  Equation\,\eqref{genThirring} reduces to the massive Thirring model for $m_2=0$. The parity operator $P$ acts as
\begin{equation}
P \psi(x, t) P=\gamma_{0} \psi(-x, t), \; P \bar{\psi}(x, t) P=\bar{\psi}(-x, t) \gamma_{0} .
\end{equation}
The time-reversal operator $T$ acts as
\begin{equation} 
T \psi(x, t) T=\gamma_{0} \psi(x,-t),  \; T \bar{\psi}(x, t) T=\bar{\psi}(x,-t) \gamma_{0},
\end{equation}
which is identical to the action of $P$ if not for the fact that $T$ is antilinear. Using these definitions, we can check that the Hamitonian is Hermitian for $m_2=0$. Note that it is also separately invariant under parity and time-reversal in this case.

The  $\gamma_{5}-$dependent mass term $m_2$ renders the Hamiltonian non-Hermitian  because the sign of the $m_{2}$-term is reverted under Hermitian conjugation. This sign change takes place because $\gamma_{0}$ and $\gamma_{5}$ anticommute. The Hamitonian is moreover not invariant under either $P$ or $T$ separately because the $m_{2}$-term changes sign under both these operations. It is however  invariant under the product operator $PT$. Thus, $H$ is PT-symmetric. 

Spontaneously broken PT symmetry manifests as the occurrence of nonreal eigenvalues for the many-body Hamiltonian\,\eqref{genThirring}. Considering the free theory ($\lambda=0$) at first, we write the field equations of motion for $\psi$, which square to a Klein-Gordon equation with square mass~\cite{BenderBook}
\begin{equation} \mu^2 =m_1^2-m_2^2. \end{equation}
The physical mass $\mu$ that propagates is therefore real for $m_1>m_2$ ($\mu^2>0$), which we refer to as the PT-unbroken sector. Conversely, $m_1<m_2$ ($\mu^2<0$) corresponds to the PT-broken sector, for which the mass is purely imaginary $\mu=\pm i|\mu|$. 

\begin{figure}[h!]
    \centering
    \includegraphics[width=0.5\textwidth]{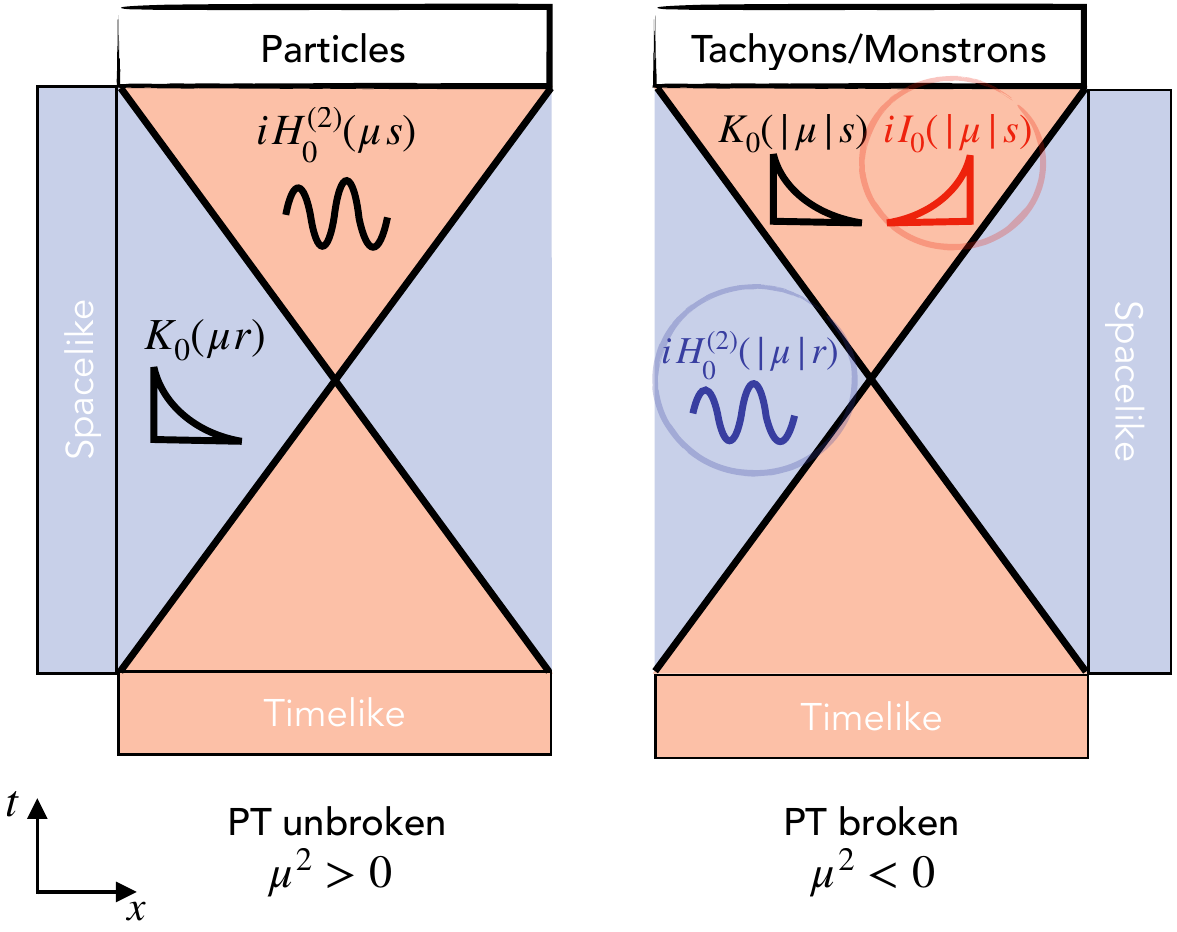}
    \caption{Schematic of the asymptotic behavior of the propagator $\Delta(x,t)$ for large spacetime intervals inside and outside of the light cone in the case of (left) unbroken ($\mu^2>0$) and (right) spontaneously broken ($\mu^2<0$) PT symmetry. For large timelike intervals (i.e. inside the light cone), while amplitudes asymptotically display oscillatory ($H_0^{(2)}$) behavior in the PT-unbroken sector, they can display exponential growth ($I_0$) in the PT-broken sector, or in other words monstronic behavior. For spacelike intervals (i.e. outside the light cone, beyond the Lieb-Robinson bound), amplitudes in the PT-broken sector asymptotically display oscillatory (tachyonic) behavior in contrast with their exponential suppression ($K_0$) when PT symmetry is unbroken. }
    \label{Tachyons}
\end{figure}

 As we will see, whether PT symmetry is spontaneously broken or not is crucial in determining the asymptotic behavior of propagators. The spacetime propagator for the generalized Thirring fermions $\Delta_F(x)$ can be written as 
\begin{equation}
    -i\Delta_F(x,t)= (i\slashed{\partial} -m_1-m_2\gamma_5)\Delta(x,t)
\end{equation}
where the form of $\Delta(x)$ is given according to the sign of $\mu^2$ as follows. For $\mu^2>0$, i.e. the  PT-symmetric regime,
\begin{equation}\label{PropagatorPTsym}
    \Delta(x,t) = \left\{\begin{array}{lll}
  -\frac{1}{2\pi} K_0(\mu r) &  \quad \mbox{(spacelike)} \\[0.5cm]
  -\frac{i}{4}H_0^{(2)}(\mu s) &  \quad \mbox{(timelike)}
\end{array}\right.  
\end{equation}  
where $ r = \sqrt{-t^2+x^2}  $ and $  s = \sqrt{+t^2-x^2} $,  $H_0^{(2)}$ is the Hankel function of the second kind which represents an outgoing wave for large times $|t|$, while $K_0$ is the modified Bessel function of the second kind displaying exponential decay at large distances $|x|$ (therefore ensuring exponential suppression of amplitudes outside of the light cone). In contrast, for $\mu^2<0$, i.e. when PT symmetry is spontaneously broken, one finds (see Appendix~\ref{TachyonPropagatorAppendix})


\begin{equation}
    \Delta(x,t)= \left\{\begin{array}{lll}
  \frac{i}{4} H_0^{(2)}(|\mu| r) &  \quad \mbox{(spacelike)}\\[0.5cm]
  \frac{1}{2\pi}K_0(|\mu| s)+ \frac{i}{2} I_0(|\mu| s) & \quad \mbox{(timelike)}
\end{array}\right. 
\end{equation} 
where $I_0$ is the modified Bessel function of the first kind, displaying exponential growth at large times $|t|$. This implies exponentially growing amplitudes for timelike intervals (i.e. within the light cone), indicating the presence of excitations that we call monstrons~\cite{Monstron_cFunction_SG}. While such unitarity-violating modes have to be excluded in closed (Hermitian) systems, nothing prevents their occurrence in  non-Hermitian settings. Outside of the light cone (beyond the Lieb-Robinson bound), we see oscillatory behavior indicative of  tachyonic modes. These results are schematically summarized in Fig.~\ref{Tachyons}. Since both types of modes are the manifestation of the imaginary mass $\mu$ associated with PT breaking, the modulus $|\mu|$ may either refer to the growth rate of the amplitudes of the monstronic modes or the oscillation rate of tachyonic amplitudes depending on whether we are inside or outside of the light cone. In the following, we shall refer for simplicity mainly to tachyons, which we use from now on as a generic term for modes characterized by an imaginary mass.\\
\vspace{-2mm}
\vspace{-6mm}
\section{Renormalization group analysis}  It is known that the Hermitian massive Thirring model admits a continuous line of interacting massless fixed points, which is either attractive or repulsive depending on the bare mass and coupling strength. We now study the behavior of the generalized massive Thirring model under renormalization to investigate how the occurrence of a $\gamma_5$-dependent mass and the interplay between interactions and non-Hermiticity modify those results. Of particular interest is the behavior of the model under renormalization when the propagating mass is imaginary in the PT-broken sector ($\mu^2<0$). The renormalization group analysis detailed in Appendix~\ref{FRG_Appendix} predicts the following flow equations for the mass $m_1$ and its $\gamma_5$-dependent counterpart $m_2$ 

\begin{equation}
\partial_{\tau} \bar{m}_i=\bar{m}_i\left[1+\frac{2 \bar{\lambda}}{\pi\left(1 + 4 \bar{\mu}^2\right)}\right],   \quad \quad (i=1,2)
\end{equation}
where  $\tau$ is the 'RG time' $|\tau|=\ln (\Lambda / k)$ and the dimensionless masses $\bar{m}_i=(2 k)^{-1} {m}_{k,i}$ and coupling strength $\bar{\lambda}=\lambda_{k} / 2$ are expressed in units of the running momentum scale $k$. The ultraviolet cutoff $\Lambda$ is naturally provided by the lattice spacing if the underlying microscopic model is defined on a lattice. 
Noticing the similarity between the flow equations for $m_1$ and $m_2$, one can show that the ratio between $m_1$ and $m_2$ remains constant under renormalization
\begin{equation}\label{Ratio}
    \partial_\tau \left(\frac{\bar{m}_1}{\bar{m}_2}\right) = 0.
\end{equation}

The structure of the RG flow  reflects the fact that the only physically-relevant mass in the problem (namely the one appearing in the propagator) is given by $\mu$. Accordingly, in the PT-unbroken sector, the Hamiltonian of the generalized non-Hermitian massive Thirring (\ref{genThirring}) is known to admit the same spectrum~\cite{BenderBook} as that of the Hermitian massive Thirring with mass $\mu = \sqrt{m_1^2-m_2^2}$. Given these considerations, the flow equations can be expressed entirely in terms of $\bar{\mu}^2=\bar{m}_1^2-\bar{m}_2^2$ (whose sign is therefore also preserved under renormalization) and $\bar{\lambda}$ as
\begin{equation}\label{FLowEq1}
    \begin{aligned}
\tfrac{1}{2}\partial_{\tau} \bar{\mu}^2&=\bar{\mu}^2\left[1+\frac{2 \bar{\lambda}}{\pi\left(1 + 4 \bar{\mu}^2\right)}\right], \\
\partial_{\tau} \bar{\lambda}&=\frac{16}{\pi} \frac{\bar{\lambda}^{2} \bar{\mu}^2}{\left(1+4 \bar{\mu}^2\right)^{2}}.
\end{aligned}
\end{equation}

\begin{figure}[h!]
    \centering
    \includegraphics[width=0.45\textwidth]{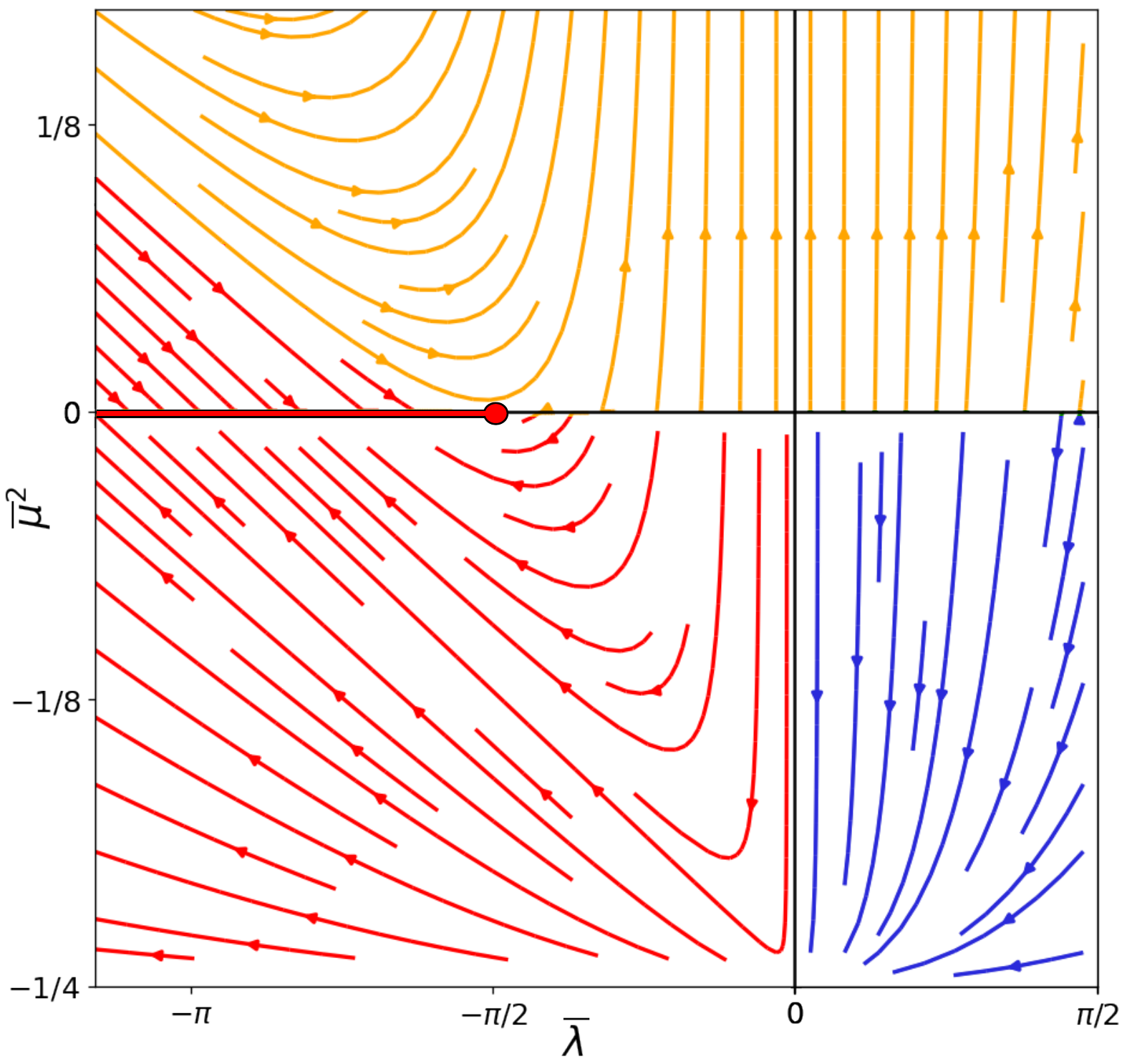}
    \caption{RG flow for the  dimensionless square mass $\bar{\mu}^2$ and quartic coupling strength $\bar{\lambda}$. In the PT-unbroken sector ($\bar{\mu}^2>0$), the flows qualitatively reproduce the traditional BKT picture: the line of fixed points with $\bar{\mu}^2=0$ that is attractive for $\bar{\lambda}\leq -\pi/2$ (red solid line) and repulsive otherwise. This results in a massless BKT (red) and a massive (yellow) phase separated by the well-known BKT point. The PT-broken sector ($\bar{\mu}^2<0$) displays for $\bar{\lambda}<0$ semicircular flows consistent with previous perturbative RG studies of the generalized sine-Gordon~\cite{Ashida_PTSG}. In this region, the BKT line is  attractive also for  $-\pi/2\leq\bar{\lambda}\leq 0$, resulting in the stabilization of the power-law scaling for all attractive interaction strengths (red). For repulsive interactions $\bar{\lambda}\geq 0$, a novel massive phase emerges  characterized by a finite imaginary mass and leading to exponentially growing (monstronic) amplitudes for large timelike intervals and oscillatory (tachyonic) amplitudes for large spacelike intervals. }
    \label{Flow_NoD}
\end{figure}

The phase diagram of the non-Hermitian Thirring model obtained from the FRG flow Eqs.\,\eqref{FLowEq1} is reported in Fig.~\ref{Flow_NoD}.
For $\bar{\mu}^2>0$ (i.e. the PT-unbroken sector), the flow reproduces the well-known BKT physics: one finds a line of fixed points at $\bar{\mu}^2=0$ that is attractive (double red line) for $\bar{\lambda}\leq -\pi/2$ and repulsive otherwise. The unbinding point at $\bar{\lambda}= -\pi/2$ is reported as a red full circle. More interestingly, in the PT-broken sector $\bar{\mu}^2<0$, the flow Eqs.\,\eqref{FLowEq1} display a spinodial line at $\bar{\mu}^2=-1/4$, where both beta functions become infinite. This spinodal line represents the UV limit of the theory. For each point above this line $\bar{\mu}^2>-1/4$ with $\bar{\lambda}<0$, the flows are semicircular and the system will eventually end up in any of the attractive BKT fixed points. In this regime, the behavior of the non-Hermitian Thirring model reproduces the one of the generalized sine-Gordon model\,\cite{Ashida_PTSG}, as would be expected from the straightforward extension of the duality\,\cite{coleman1975quantum, Duality} to the PT-broken phase. 

This analogy fails when the fermions, representing the sine-Gordon solitons, become repulsive, i.e. when $\bar{\lambda}>0$. While the PT-unbroken sector displays the same infrared behavior for any $\bar{\lambda}> -\pi/2$ (yellow region in Fig.~\ref{Flow_NoD}), a novel infrared phase emerges in the PT-broken sector of the massive Thirring model for $\bar{\lambda}>0$ (blue region in Fig.~\ref{Flow_NoD}).  There, the flow is attracted to an infrared spinodal point $(\bar{\mu}^2,\bar{\lambda})=(-1/4,0)$, which is the termination of the UV spinodal line described for $\bar{\lambda}<0$. This point  corresponds to a noninteracting theory with an imaginary propagating mass ($\bar{\mu}^2=-1/4$). Yet, the spinodal nature of this attractive infrared point causes the flow to terminate at a finite scale $k_{c}$ in its vicinity\,\cite{IRFP_Spinodial}. { The spinodal point stemming from the singularity in the beta functions is also known to arise in Hermitian systems. In Hermitian systems, $k_{c}$  represents  a finite correlation length scale of the system and the value of the couplings at the breakdown of the flow coincide with the thermodynamic properties of the system in the massive infrared phase\,\cite{IRFP_Spinodial}.}  These effective parameters describing the renormalized theory are nonuniversal and depend on the microscopic (bare) initial conditions of the flow.  Therefore, the spinodal infrared point describes a noncritical massive phase.

At $\bar{\lambda}=0$, a phase transition occurs between a massless (interacting) phase and a phase with an imaginary propagating mass.  As explained above, this phase is characterized by tachyonic modes. Since the flows { interrupt} before reaching the singularity, the renormalized (imaginary) mass (meaning the mass obtained at the end of the flow) depends on the microscopic parameters of the model, providing a way to tune the corresponding mode oscillation/amplification  rate.  {We mention that an infrared phase was evidenced in the nonperturbative treatment of the non-Hermitian sine-Gordon model. There, the semicircular flows occurring in the PT-broken phase break down at a finite value of the superfluid stiffness $K$, where a new infrared phase appears\,\cite{AshidaThesis}.  Whereas this phase has found no clear interpretation in the sine-Gordon paradigm, the fermionic Thirring picture provides a physical meaning for this phase as a truly non-Hermitian phase hosting tachyonic excitations occurring when the interaction between the fermions (the solitons of the sine-Gordon model~\cite{Coleman_Duality}) becomes repulsive in the PT-broken sector. In the following we will show how the transparency of the Thirring picture allows us to generalize our findings to the presence of external gauge fields.} 



\section{Generalized massive Thirring with external gauge field}

We now study the influence of an additional external gauge field on the phase diagram of the generalized massive Thirring model, with particular focus on the interplay between such interactions and spontaneous symmetry breaking of PT symmetry. As mentioned before, these terms may be used to model various phenomena depending on the setting such as incommensurabilities in classical systems~\cite{Incomm, Pert_RG_Incomm}, magnetic fields in spin systems~\cite{ChitraSpinB} soliton fugacities~\cite{Soliton_density} in bosonic settings or chemical potentials~\cite{Schutz_LatticeFermions} in fermionic systems. The modified Hamiltonian obtained  via the minimal substitution $\partial_\mu\rightarrow D_\mu = \partial_\mu - id_\mu $ with $d_0=d$ and $d_1=0$ reads
\begin{equation}\label{genThirringGauge}
    H=\int dx \left[\bar{\psi}\left(-i\slashed{\nabla}+d\gamma_0+  {m}_1+{m}_2 \gamma_5\right) \psi+\lambda (\bar{\psi} \psi)^{2}\right].
\end{equation}

\begin{figure}[h!]
    \centering
    \includegraphics[width=0.45\textwidth]{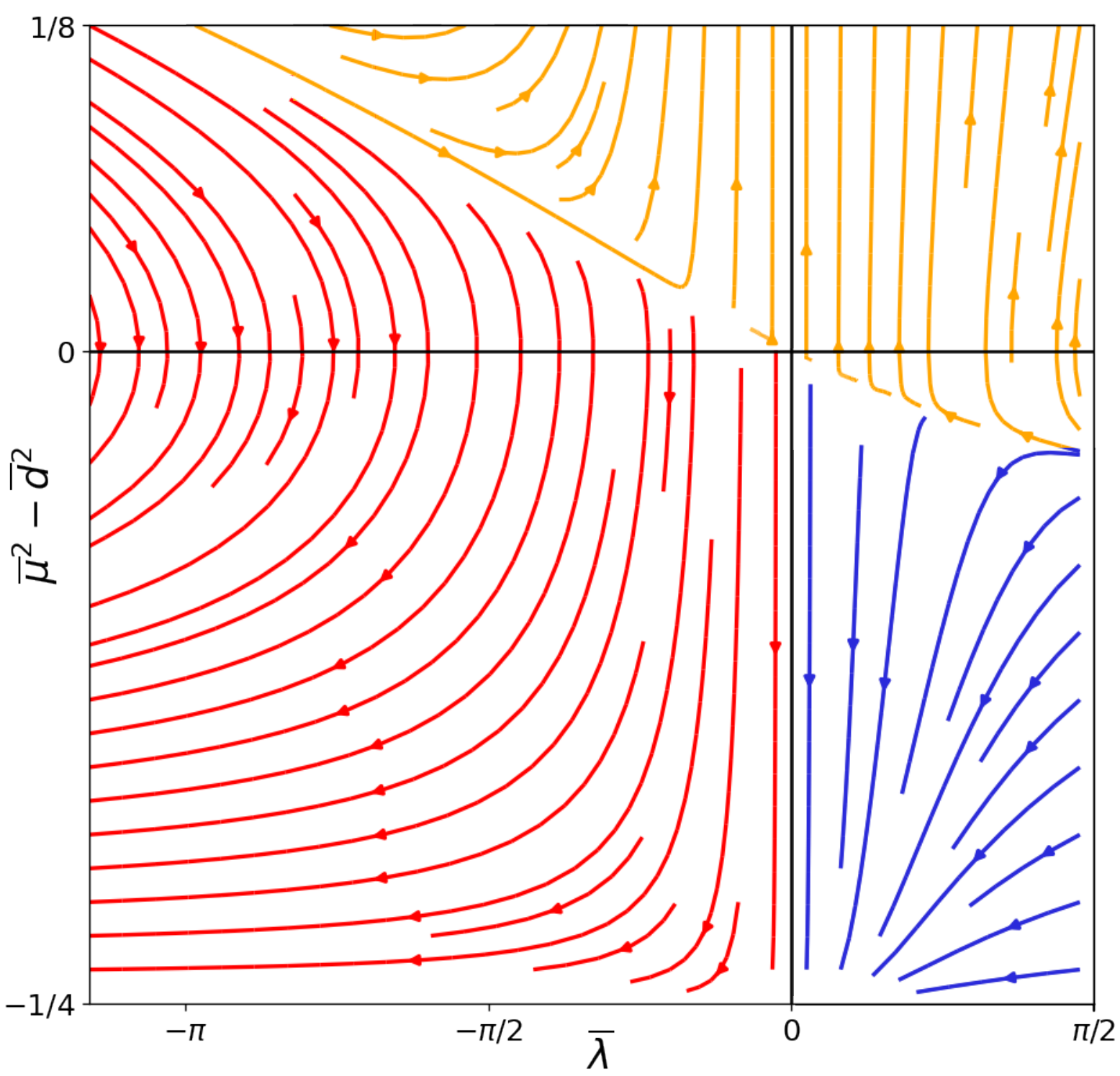}
    \caption{RG flow slice for the coupling combination $\bar{\mu}^2-\bar{d}^2$ and the quartic coupling strength $\bar{\lambda}$ (for $\bar{d}$ fixed at $0.3$, other values of $\bar{d}$ lead to qualitatively equivalent flow charts). In particular, the flow  no longer displays $\bar{\mu}^2=0$ as a continuum of fixed points for $\bar{d}\neq 0$, effectively ruling out the BKT transition as pointed out in a recent study~\cite{Russian_FRGThirring}. For points belonging to a region (in blue) included in the lower-left quadrant ($\bar{\lambda}>0$ and $\bar{\mu}^2<d^2$), the flows converge to $\bar{\mu}^2-\bar{d}^2 = -1/4$  and $\bar{\lambda} = 0$. This includes the PT-broken sector for which $\bar{\mu}^2<0$ and a part of of the PT-unbroken sector $\bar{\mu}^2>0$.}
    \label{Flow_MU2D2}
\end{figure}

  We first note that $\mu^2=m_1^2-m_2^2$ remains the effective mass even when $d\ne 0$ (this can be checked by writing the field equations of motion for the gauge-transformed fermions   $\tilde{\psi}\equiv e^{id^\mu x_\mu} \psi$). 
The RG flow reflects this fact  and  (\ref{Ratio}) remains unchanged.  The  associated  flow equations derived in Appendix~\ref{FRG_Appendix} can again be expressed in terms of $\bar{\mu}^2$
\begin{equation}\label{FlowGauge}
    \begin{aligned}
\tfrac{1}{2}\partial_{\tau} \bar{\mu}^2&=\bar{\mu}^2\left(1+\frac{\bar{\lambda}}{4 \pi \bar{d}^{2}} \frac{\left(1-4 \bar{d}^{2}+4 \bar{\mu}^2\right)}{\sqrt{\left[2\left(\frac{1-4 \bar{d}^{2}+4\bar{\mu}^2}{4 \bar{d}}\right)^{2}+1\right]^{2}-1}}\right),\\ 
\tfrac{1}{2}\partial_{\tau} \bar{d}^2&=\bar{d}^2-\frac{\bar{\lambda}}{4 \pi} \\+&\frac{\bar{\lambda}}{4 \pi} \frac{\left(1-4 \bar{d}^{2}+4 \bar{\mu}^2\right)\left(1+\frac{1}{8 \bar{d}^{2}}\left[1-4 \bar{d}^{2}+4 \bar{\mu}^2\right]\right)}{\sqrt{\left[2\left(\frac{1-4 \bar{d}^{2}+4 \bar{\mu}^2}{4 \bar{d}}\right)^{2}+1\right]^{2}-1}}, \\
\partial_{\tau} \bar{\lambda}&=-\frac{2 \bar{\lambda}^{2}}{\pi}\left(1-\frac{\bar{\mu}^2}{\bar{d}^{2}}\right) \frac{1}{\sqrt{\left[2\left(\frac{1-4 \bar{d}^{2}+4 \bar{\mu}^2}{4 \bar{d}}\right)^{2}+1\right]^{2}-1}}.
\end{aligned}
\end{equation}
The flow diagram for the following combination of couplings $(\bar{\mu}^2-\bar{d}^2, \bar{\lambda})$ is displayed in Fig.~\ref{Flow_MU2D2}. 

Comparing Figs.~\ref{Flow_NoD} and~\ref{Flow_MU2D2}, we see that the flow lines are strongly modified by the gauge field. As a notable feature, for $d\neq0$, the flow equations no longer display $\bar{\mu}^2=0$ as a continuum line of fixed points (parametrized by the interaction $\bar{\lambda}$), effectively ruling out the BKT transition. This is consistent with a recent study focusing on the Hermitian model~\cite{Russian_FRGThirring}. Instead, the trajectories in red in Fig.~\ref{Flow_MU2D2} flow towards increasingly strong attractive interactions $\bar{\lambda}<0$. 
 Note that for  $\bar{\lambda} \ll 0$, the first equation in~\eqref{FlowGauge} approximates at leading order in $\bar{\lambda}$ to
\begin{equation} \tfrac{1}{2}\partial_{\tau} \bar{\mu}^2 = \bar{\mu}^2\bar{\lambda}f(\bar{\mu},\bar{d})
\end{equation} 
where $f$ is a positive function of $\bar{\mu}$ and $\bar{d}$ in the range of parameter values considered. In particular, this means that $\bar{\mu}^2$ and the corresponding beta function $\partial_\tau\bar{\mu}^2$ have opposite signs. In other words, the rate of change in the value of $\bar{\mu}^2$ along the flow ($\tau$) is positive when $\bar{\mu}^2$ is negative and conversely the rate of change is negative when $\bar{\mu}^2$ is positive. Therefore, asymptotically, the value of $\bar{\mu}^2$ is driven to zero and the  trajectories in the lower-left sector of the phase diagram in Fig.~\ref{Flow_MU2D2} converge towards $\bar{\mu}^2=0$. { Thus, while the original gapless BKT phase (in red in Fig.~\ref{Flow_NoD}) is lost in the presence of external gauge fields, the corresponding phase at finite $d$ (in red in Fig.~\ref{Flow_MU2D2}) remains massless nonetheless. }

\begin{figure}[h!]
    \centering
\includegraphics[width=0.5\textwidth]{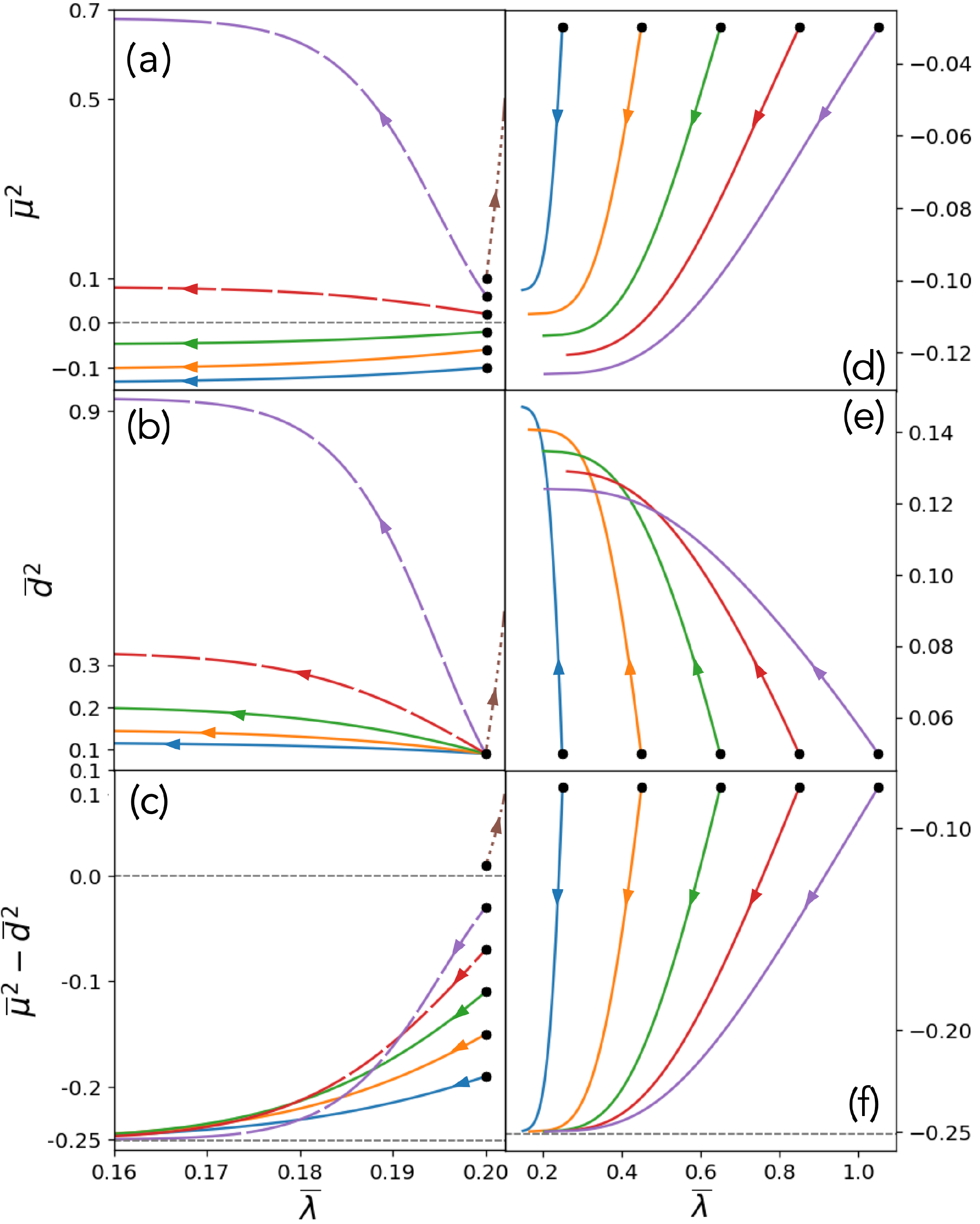}
     \caption{Flow diagrams for the dimensionless propagating mass $\bar{\mu}^2$, external gauge field $\bar{d}^2$ as well as the combination $\bar{\mu}^2 -\bar{d}^2$ together with $\bar{\lambda}$ for different initial conditions obtained by varying the bare value of the mass $\bar{\mu}_0^2$ in panels (a), (b) and (c) or by tuning the strength of the bare interaction $\bar{\lambda}_0$ in the region of positive $\bar{\lambda}$ in panels (d), (e) and (f). Unless the initial value for $\bar{\mu}^2 -\bar{d}^2$ is above a given threshold (dotted trajectories), $\bar{\lambda}$ is irrelevant and $\bar{\mu}^2 -\bar{d}^2$ converges towards $-1/4$. For all trajectories in this regime, one observes that different bare/initial conditions lead to different values of the renormalized/final square mass $\bar{\mu}^2$, to which $\bar{d}^2$ adapts since the renormalized/final difference is constrained. This is the case for the region of the PT-unbroken ($\bar{\mu}^2>0$) sector in the aforementioned regime (dashed lines) and for the PT-broken sector (solid lines). The renormalized propagating mass can therefore be tuned through the microscopic (bare) parameters of the model.} 
    \label{Tuning}
\end{figure}

As shown in Fig.~\ref{Flow_MU2D2}, the point $(\bar{\mu}^2-\bar{d}^2, \bar{\lambda}) = (-1/4,0)$ has a basin of attraction (in blue) contained in the region defined by ${\bar\lambda}>0$ and $\bar{\mu}^2<\bar{d}^2$ (the lower-right quadrant).  $\bar{\mu}^2-\bar{d}^2=-1/4$ allows for two different possibilities: either $\bar{\mu}^2=\bar{d}^2-1/4>0$ and we are in the PT-symmetric phase or $\bar{\mu}^2=\bar{d}^2-1/4<0$ and PT-symmetry is spontaneously broken. The attractive point $(\bar{\mu}^2-\bar{d}^2, \bar{\lambda}) = (-1/4,0)$ can therefore still characterize a PT-broken phase.  Hence, in contrast  to the BKT phase which is wiped out by the gauge field, the new tachyonic phase is preserved. While bare couplings $\bar{d}_0$, $\bar{\lambda}_0$ may take any value in $(\bar{\mu}_0^2,\bar{d}^2_0)\in \mathbb{R}\times \mathbb{R}^+$, the asymptotic (renormalized) values they reach at the end of the flow in the phase in blue in Fig.~\ref{Flow_MU2D2} are subject to the constraint $\bar{\mu}^2- \bar{d}^2=-1/4$. This leaves in particular one degree of freedom: as we will see, in this part of the phase diagram, the renormalized mass itself (i.e. the value of $\bar{\mu}^2$ reached at the end of the flow) can be continuously tuned via the bare parameters $\bar{\mu}_0$, $\bar{d}_0$ and $\bar{\lambda}_0$. This is demonstrated in
Fig.~\ref{Tuning}  which displays the flow of $\bar{\mu}^2$, $\bar{d}^2$ and $\bar{\mu}^2-\bar{d}^2$ together with $\bar{\lambda}$ for different starting points (solid black dots) corresponding to distinct bare values of the coupling strengths. 

We first focus on the panels (a), (b) and (c) in Fig.~\ref{Tuning}, which  displays  the flows of  $\bar{\mu}^2$, $\bar{d}^2$ and  $\bar{\mu}^2-\bar{d}^2$, respectively, as a function of $\bar{\lambda}$ for different initial (bare) values of the propagating mass $\bar{\mu}_0$. The bare values  of the gauge field $\bar{d}_0$ and  coupling  $\bar{\lambda}_0$ are kept fixed. Three different regimes are identified, corresponding respectively to the dotted, dashed and solid flow lines. In panel (c), there exists a threshold value $\bar{\mu}^2_0-\bar{d}^2_0\approx 0$ for the bare difference in coupling strengths above which both $\bar{\lambda}$ and $\bar{\mu}^2-\bar{d}^2$ flow towards arbitrary high values (dotted curve). In this regime, both $\bar{\mu}$ and $\bar{d}$ diverge (see panels (a) and (b)). This phase corresponds to the yellow region in the upper-right ($\bar{\lambda}>0$) quadrant in Fig.~\ref{Flow_MU2D2}. In contrast, below this threshold,  we have flows which converge  to  $\bar{\mu}^2-\bar{d}^2=-1/4$ and $\bar{\lambda} = 0$  (dashed and solid lines in panel (c)). This corresponds to the blue region in the lower-right quadrant ($\bar{\lambda}>0$) in Fig.~\ref{Flow_MU2D2}.  From panel (a), we infer that these trajectories  arise in both the unbroken PT sector (dashed lines) where $\bar{\mu}^2>0$ and the PT-broken sector (solid lines) where $\bar{\mu}^2<0$.  In both cases, each value of the bare mass $\bar{\mu}_0^2$ (solid black dots) leads to a different value of the renormalized mass $\bar{\mu}^2$. The value of $\bar{d}^2$ for the dashed and solid trajectories adapts correspondingly, see panel (b), so that  $\bar{\mu}^2-\bar{d}^2\to -1/4$ as displayed in panel (c).

Panels (d), (e) and (f)  show  the flows of  $\bar{\mu}^2$,  $\bar{d}^2$ and  $\bar{\mu}^2-\bar{d}^2$ as a function of  $\bar{\lambda}$ for different initial (bare) values $\bar{\lambda}_0$ of the quartic coupling strength.  The bare values  $\bar{d}_0,\bar{\mu}_0$ are kept fixed and chosen in the PT-broken sector ($\bar{\mu}^2<0$) so that the trajectories belong to the blue region in Fig.~\ref{Tuning}. As shown in panel (d), each value  of the bare coupling strength $\bar{\lambda}_0$ (solid black dots) leads to a different value of the renormalized mass $\bar{\mu}^2$. Once again, $\bar{d}$ adapts, see panel (e), so that $\bar{\mu}^2-\bar{d}^2\to -1/4$ asymptotically as shown in panel (f).
To summarize, we see
that the renormalized mass and the corresponding asymptotic properties of the propagator can be tuned via the bare parameters of the model ($\bar{\mu}_0$, $\bar{d}_0$, $\bar{\lambda}_0$). In particular, one can tune in the PT-broken sector the amplification/oscillation rate of amplitudes of the tachyonic modes. We note that this applies to a part of the PT-unbroken sector in which the (now real) mass of the particle modes can be tuned as well.\\


\section{Conclusion/Discussion}

The functional renormalization group was employed to  investigate putative  phase transitions and the influence of external gauge fields in a generalized PT-symmetric two-dimensional massive Thirring model. We see that this model spontaneously breaks PT symmetry in a certain coupling regime.  We find that for attractive interactions, PT symmetry is restored in the asymptotic flow.  However,  for repulsive interactions,  our analysis predicts a novel  PT-broken phase, which  manifests  as an attractive spinodal fixed point of the flow equations, characterized  by the termination of the flow at a finite nonzero momentum $k_{c}$.  This phase harbors an imaginary mass describing tachyonic\, excitations, which  display exponentially\,(oscillatory) amplitudes for timelike\,(spacelike) intervals and potentially violate  Lieb-Robinson bounds.  

The presence of an additional gauge field is shown to dramatically alter  the renormalization group flows. The tachyonic PT-broken phase is however robust to this  gauge field.  As opposed to standard massive phases, the renormalized imaginary mass of the tachyons can be tuned in a controlled manner via the microscopic parameters. 

Future perspectives include the quantitative study of the relation between the generalized Thirring model and the PT-broken sine-Gordon model\,\cite{Ashida_PTSG}, generalizing the duality argument\,\cite{coleman1975quantum, Duality} to the PT-broken setting. More generally, it is of great interest to investigate fundamental questions which remain unanswered regarding the consistent path-integral description of PT-broken phases. Note that the results of renormalization group procedures should nevertheless be independent of the description and have been applied (together with other procedures such as the thermodynamic Bethe ansatz, finite size analysis or scattering theory) directly onto PT-broken action functionals while showing consistency with numerical simulations~\cite{Ashida_PTSG, Monstron_cFunction_SG}. Other possible enlightening perspectives might include the computation of the mass of lowest excitation as a function of the interaction strength in the PT-symmetric Thirring/sine-Gordon models, aiming to generalize previous results based on the derivative expansion in the Hermitian setting~\cite{Dupuis_FRG_Spectrum} to the PT-broken regime.  Lastly, it would be exciting to explore prospects for realizing such novel  monstronic/tachyonic modes stemming from PT symmetry breaking  in realistic experimental platforms  which can simulate non-Hermitian Hamiltonians such as ultracold atoms and exciton-polariton condensates~\cite{RealisePTQMBS, Exp_2, Exp_3, Exp_5, Exp_6, Ref_PseudoYL_Exp1, Ref_PseudoYL_Exp2,Ref_PseudoYL_Exp3,Ref_PseudoYL_Exp4}.

\acknowledgments

This work was supported by an ETH Zürich Research Grant and the Deutsche Forschungsgemeinschaft (DFG, German Research Foundation) under Germany’s Excellence Strategy EXC2181/1-390900948 (the Heidelberg STRUCTURES Excellence Cluster).

\vspace{1.5cm}

\appendix
\onecolumngrid
\appendix
\section{Tachyonic propagator in the PT-broken phase} \label{TachyonPropagatorAppendix}

Whether PT symmetry is spontaneously broken (or equivalently the sign of the square propagating mass $\mu^2$) is crucial to determine the asymptotic decay of propagators. A standard calculation in the PT-symmetric case ($\mu^2>0$) leads to asymptotic properties of amplitudes analogous to the purely Hermitian case: Eq.~(\ref{PropagatorPTsym}) only differs from the Hermitian case through the substitution $\mu\rightarrow m_1$. In contrast, we will see that PT symmetry breaking ($\mu^2 < 0$) has dramatic consequences for the asymptotic decay of propagators.  The fermion propagator can be generally obtained as follows

\begin{equation}
    -i\Delta_F(x,t) = (i\slashed{\partial}-m_1-m_2 \gamma_5)\Delta(x,t), \mbox{ where } \Delta(x,t)\equiv \int \frac{d^{2} k}{(2 \pi)^{2}} \frac{i e^{-i k^\mu x_\mu}}{k^{2}-\mu^{2}+i \epsilon},
\end{equation}

in which $\epsilon\rightarrow 0^+$. In the PT-broken phase, the propagating mass $\mu$ becomes imaginary $\mu=i|\mu|$. We now derive a closed form for $\Delta(x,t)$ in this particular case:

\begin{equation}
\begin{aligned}
\Delta(x,t)&=\int \frac{d^{2} k}{(2 \pi)^{2}} \frac{i e^{-i k^\mu x_\mu}}{k^{2}+|\mu|^{2}+i \epsilon} \\
&= \int \frac{d k_1 d k_{0}}{(2 \pi)^{2}} \frac{i e^{ -i k_{0} t+i {k}_1{x}}}{k^{2}_0-k_1^2+|\mu|^{2}+i \epsilon} \\
&=i\int_{|{k}_1| < |\mu|} \frac{d {k}_1}{(2 \pi)} \frac{e^{- \sqrt{|\mu|^{2}-k_1^2}|t|}}{2 \sqrt{|\mu|^{2}-k_1^2}}e^{+i {k}_1{x}} +\int_{|{k}_1| > |\mu|} \frac{d {k}_1}{(2 \pi) }\frac{e^{-i \sqrt{k_1^2-|\mu|^{2}}|t|}}{2 \sqrt{k_1^2-|\mu|^{2}}}e^{+i {k}_1{x}}\\
&=i\int_{0}^{|\mu|} \frac{d {k}_1}{(2 \pi)  } \frac{e^{- \sqrt{|\mu|^{2}-k_1^2}|t|}}{\sqrt{|\mu|^{2}-k_1^2}} \cos(k_1x) +\int_{|\mu|}^\infty \frac{d {k}_1}{(2 \pi)}\frac{e^{-i \sqrt{k_1^2-|\mu|^{2}}|t|}}{  \sqrt{k_1^2-|\mu|^{2}}}\cos(k_1x),
\end{aligned}
\end{equation}
in which we have evaluated the integral in $k_0$ as follows
\begin{equation}
\hspace{-2mm} \int_{-\infty}^{+\infty} \frac{d k_{0}}{2\pi} \frac{i e^{-i k_{0} t}}{{k^{2}_0-k_1^2+|\mu|^{2}+i \epsilon}} = \frac{e^{-i \sqrt{k_1^2-|\mu|^{2}}|t|}}{2\sqrt{k_1^2-|\mu|^{2}}}\theta(k_1^2-|\mu|^2) + \frac{e^{-\sqrt{|\mu|^{2}-k_1^2}|t|}}{-2i\sqrt{|\mu|^{2}-k_1^2}}\theta(|\mu|^2-k_1^2), 
\end{equation}
taking into account the fact that for $|k_1|>|\mu|$, the poles are on the real axis at $k_0 = \pm\sqrt{k_1^2-|\mu|^2} $ while for $|k_1|<|\mu|$, they are on the imaginary axis at $k_0 = \pm i \sqrt{|\mu|^2-k_1^2} $.
For spacelike intervals, we can pick a frame of reference such that the spacetime interval is purely spatial and revert back to the original coordinates to find

\begin{equation}\begin{aligned}
\Delta(x,t) &= i\int_{0}^{|\mu|} \frac{d {k}_1}{(2 \pi)  } \frac{\cos(k_1r)}{\sqrt{|\mu|^{2}-k_1^2}}  +\int_{|\mu|}^\infty \frac{d {k}_1}{(2 \pi)}\frac{\cos(k_1r)}{  \sqrt{k_1^2-|\mu|^{2}}}\\
&=  \frac{i}{4}J_0(|\mu|r)-\frac{1}{4}Y_0(|\mu|r)\\
&=  \frac{i}{4} H_0(|\mu|r),
\end{aligned}\end{equation}
where $ r = \sqrt{-t^2+x^2}$. Similarly for timelike intervals, we can write
\begin{equation}
\begin{aligned}
\Delta(x,t) &=i\int_{0}^{|\mu|} \frac{d {k}_1}{(2 \pi)  } \frac{e^{- \sqrt{|\mu|^{2}-k_1^2}s}}{\sqrt{|\mu|^{2}-k_1^2}}  +\int_{{|\mu|}}^\infty \frac{d {k}_1}{(2 \pi)}\frac{e^{-i \sqrt{k_1^2-|\mu|^{2}}s}}{  \sqrt{k_1^2-|\mu|^{2}}}\\
&=i\int_{0}^{|\mu|} \frac{d p}{(2 \pi) } \frac{e^{- ps}}{\sqrt{|\mu|^{2}-p^2}}  +\int_{{|\mu|}}^\infty \frac{dl}{(2 \pi)}\frac{e^{-i ls}}{  \sqrt{l^2+|\mu|^{2}}}\\
&=  \frac{i}{2} I_0(|\mu|s)+\frac{1}{2\pi} K_0(|\mu| s),
\end{aligned}    
\end{equation}
where $ s = \sqrt{+t^2-x^2}$. Note that the fact that the presence of the Bessel function $I_0(|\mu|s)$ implies exponential growth of amplitudes with time. This intrinsically violates unitarity of dynamics, which is generally assumed. This apparent problem is sometimes cured through the instauration of an infrared cutoff for $k_1$ at $|\mu|$~\cite{TachyonsProp}. In the non-Hermitian setting, these cannot however be excluded, particularly in the PT-broken phase where unitarity cannot be restored through a choice of scalar product~\cite{BenderBook}.\\

Note the qualitative difference in the pole structure of the momentum-space propagators in the PT-symmetric and PT-broken phases: 
\begin{equation}
\Delta(k_0,k_1) = \begin{cases}
        \displaystyle \frac{1}{k^{2}_0-k_1^2-\mu^{2}+i \epsilon} & \displaystyle\mu^2>0\\
        \displaystyle\frac{1}{k^{2}_0-k_1^2+|\mu|^2+i \epsilon} & \displaystyle \mu^2<0\\
\end{cases}.
\end{equation}
The poles in the PT-symmetric phase are given in the complex $k_0$-plane by $(-\infty,-\mu]\cup[+\mu,+\infty)$ while they are given by $\mathbb{R}\cup i[-|\mu|,|\mu|]$ in the PT-broken phase. In particular, at the exceptional point ($\mu^2=0$), the spectrum transitions from being fully real to being partially imaginary. The presence of this singular point marking the qualitative difference between the PT-symmetric and PT-broken phase prevents the use of classical analytical continuation tools such as identities for Bessel and Hankel functions for imaginary arguments that would connect the propagators in the two phases.

\section{Derivation of the flow equations}\label{FRG_Appendix}

 The flow equations~(\ref{FlowGauge}) are derived from the Wetterich equation based on a LPA scheme coupled with an expansion in the field-dependent part of the inverse regularized propagator applied onto the euclidean action corresponding to the generalized massive Thirring Hamiltonian in the presence of an external gauge field, see Eq.\,\eqref{genThirringGauge}. Equations\,\eqref{FlowGauge} reduce to Eq.\,\eqref{FLowEq1} for $d=0$. The sharp cutoff $
r\left(\frac{q^{2}}{k^{2}}\right)=\left\{\begin{array}{cc}
\infty, & q^{2}<k^{2} \\
0, & q^{2}>k^{2}
\end{array}\right.
$ is employed, which facilitates explicit evaluation of the threshold functions. Details of the calculation are provided below and closely follow the FRG treatment of the Hermitian massive Thirring model in the presence of external gauge fields~\cite{Russian_FRGThirring} while generalizing them to the non-Hermitian case. In particular, the computation differs in the additional presence of a $\gamma_5$-dependent mass term $\gamma_5 m_2$ with ${\gamma_5=\sigma_3}$.

 We start by writing the Euclidean action corresponding to the Hamiltonian (\ref{genThirringGauge}) for the two-dimensional massive Thirring model in the presence of external gauge fields 

\begin{equation}\label{EuclideanAction}
    S=i\int d^{2} r\left[\bar{\psi}\left( \sigma_{\mu} D_{\mu}+ m_1+m_2 \sigma_3\right) \psi+\lambda (\bar{\psi} \psi)^{2}\right],
\end{equation}
where the covariant derivative $D_{\mu}=\partial_{\mu}-d \delta_{\mu, 1}$  is given in terms of the external gauge field $d$. Through a Fourier transform
\begin{equation}\label{Fourier}
\psi(\mathbf{x})=\int \frac{d^{2} \mathbf{p}}{(2 \pi)^{2}} e^{i \mathbf{p x}} \psi_{\mathbf{p}} \equiv \int_{\mathbf{p}} e^{i \mathbf{p x}} \psi_{\mathbf{p}},
\end{equation}
the action~(\ref{EuclideanAction}) can be written as follows in momentum space:
\begin{equation}\label{ActionMomentum}
\begin{aligned}
{S} =& \int \frac{d^{2} \mathbf{p}}{(2 \pi)^{2}} \bar{\psi}_{\mathbf{p}}(-\slashed p-i \slashed{d}+i {m}_1+im_2\sigma_3) \psi_{\mathbf{p}}+(2 \pi)^{2}\lambda \\
& \times \int \prod_{i=1}^{4} \frac{d^{2} p_{i}}{(2 \pi)^{2}}\left(\bar{\psi}_{\mathbf{p}_{1}} \psi_{\mathbf{p}_{2}}\right)\left(\bar{\psi}_{\mathbf{p}_{3}} \psi_{\mathbf{p}_{4}}\right) \\
& \times \delta\left(-\mathbf{p}_{{1}}+\mathbf{p}_{{2}}-\mathbf{p}_{{3}}+\mathbf{p}_{{4}}\right),
\end{aligned}
\end{equation}
where the (newly redefined) slashed notations $\slashed{p}=p_{\mu} \sigma_{\mu}, \slashed{d}=d \sigma_{1}$ are used.

  Wetterich's formulation of RG is based on the effective average action $\Gamma_{k}$,  a generalization of the effective action retaining only rapidly oscillating modes, namely fluctuations with wave vectors satisfying $q^{2} \geqslant k^{2}$, where $k$ is a UV cutoff for slowly varying modes~\cite{54}. This is implemented through the use of a regulator (IR cutoff) $R_{k}$ in the inverse propagator. Slowly oscillating modes with momenta $q^{2} \leqslant k^{2}$ are decoupled through the regulator which largely increases their mass, while leaving high-momentum modes unaffected.

The Wetterich equation governs how $\Gamma_{k}$ depends on the scale $k$
\begin{equation} \label{Wetterich}
\partial_{k} \Gamma_{k}=-\frac{1}{2} \operatorname{Tr}\left[\frac{\partial_{k} R_{k}}{\Gamma^{(2)}+R_{k}}\right]=-\frac{1}{2} \tilde{\partial}_{k} \operatorname{Tr} \log \left(\Gamma^{(2)}+R_{k}\right),
\end{equation}
with $\Gamma^{(2)}$ denoting the second derivative of $\Gamma_{k}$ with respect to the fields. The trace is here meant as both an integration over momenta as well as a sum over internal indices. The tilde on the derivative $\tilde{\partial}_{k}$ is intended as a sign that it acts only on the $k$-dependence of the regulator $R_{k}$ and not on that of $\Gamma^{(2)}$. The negative sign on the right-hand side of Eq.~(\ref{Wetterich}) is a consequence the Grassman nature of the field $\psi$~\cite{55}.

The average action at $k=0$ is by definition  identical to the effective action, as the IR cutoff is not present and all modes are included. Similarly to~(\ref{ActionMomentum}), we make the following ansatz for the average action 
\begin{equation}\label{EffectiveAction}
\begin{aligned}
\Gamma_{k}=& \int \frac{d^{2} \mathbf{p}}{(2 \pi)^{2}} \bar{\psi}_{\mathbf{p}}\left(- \slashed p-i \slashed d_{k}+i {m}_{1,k}+i{m}_{2,k}\sigma_3\right) \psi_{\mathbf{p}}+(2 \pi)^{2} \\
& \times\lambda_k \int \prod_{i=1}^{4} \frac{d^{2} \mathbf{p}_{i}}{(2 \pi)^{2}}\left(\bar{\psi}_{\mathbf{p}_{1}} \psi_{\mathbf{p}_{2}}\right)\left(\bar{\psi}_{\mathbf{p}_{3}} \psi_{\mathbf{p}_{4}}\right) \\
& \times \delta\left(-\mathbf{p}_{1}+\mathbf{p}_{2}-\mathbf{p}_{3}+\mathbf{p}_{4}\right),
\end{aligned}
\end{equation}
where the momentum-scale index $k$ indicates the scale-dependence of all parameters in the effective action.

  Using Eq.\,\eqref{Wetterich}, the effects of the interaction $\lambda$ on the various parameters and in particular their flows and fixed points can be studied systematically. To be able to make analytical predictions, one may decompose the inverse regularized propagator into a field-independent $\left(\Gamma_{k, 0}^{(2)}+R_{k}\right)$ part and a field-dependent $\left(\Delta \Gamma_{k}^{(2)}\right)$ part, which gives
\begin{equation}\label{ExpansionInverseRegPropagator}
\begin{aligned}
\frac{\partial_{k} R_{k}}{\Gamma^{(2)}+R_{k}}=& \tilde{\partial}_{k} \log \left(\Gamma_{k, 0}^{(2)}+\Delta \Gamma_{k}^{(2)}+R_{k}\right) \\
=& \tilde{\partial}_{k} \log \left(\Gamma_{k, 0}^{(2)}+R_{k}\right)+\tilde{\partial}_{k} \frac{\Delta \Gamma_{k}^{(2)}}{\Gamma_{k, 0}^{(2)}+R_{k}} \\
&-\frac{1}{2} \tilde{\partial}_{k}\left(\frac{\Delta \Gamma_{k}^{(2)}}{\Gamma_{k, 0}^{(2)}+R_{k}}\right)^{2}+\cdots
\end{aligned}
\end{equation}
The regulator may be selected as follows~\cite{48,63}
\begin{equation}\label{Regulator}
R_{k}=-\frac{\delta_{p, q}}{(2 \pi)^{2}}  r\left(\frac{q^{2}}{k^{2}}\right)\left(\begin{array}{cc}
0 & \slashed q^{T} \\
 \slashed q & 0
\end{array}\right)
\end{equation}
for Thirring fermions. We now compute explicitly the field-independent part of the inverse (regularized) propagator and the corresponding field-dependent part.\\


 The second derivatives of $\Gamma_{k}$ with respect to the fields $\psi$ and $\bar{\psi}$ can be obtained from Eq.~(\ref{EffectiveAction})
\begin{equation}\label{SecondDerivatives}
\Gamma_{k}^{(2)}(p, q)=\left(\begin{array}{ll}
\vec{\partial}_{\psi_{-p}^{T}} \Gamma_{k} \overleftarrow{\partial}_{\psi_{q}} & \vec{\partial}_{\psi_{-p}^{T}} \Gamma_{k} \overleftarrow{\partial}_{\bar{\psi}_{-q}^{T}} \\
\vec{\partial}_{\bar{\psi}_{p}} \Gamma_{k} \overleftarrow{\partial}_{\psi_{q}} & \vec{\partial}_{\bar{\psi}_{p}} \Gamma_{k} \overleftarrow{\partial}_{\bar{\psi}_{-q}^{T}}
\end{array}\right), 
\end{equation}
which yields the field-independent part
\begin{equation}\label{FieldIndependentDerivatives}
\Gamma_{k, 0}^{(2)}(p, q)=\frac{\delta_{p, q}}{(2 \pi)^{2}}\left(\begin{array}{cc}
0 & - \slashed p^{T}+i \slashed{d}^{T}_k-i {m}_{1,k} {-im_{2,k}\sigma_3} \\
- \slashed{p}-i  \slashed{d}_k+i {m}_{1,k}{+im_{2,k} \sigma_3} & 0
\end{array}\right) .
\end{equation}
The corresponding inverted regularized propagator therefore reads
\begin{equation}\label{InvertedRegProp}
\begin{aligned}
&\hspace{-6mm}\left(\Gamma_{k, 0}^{(2)}+R_{k}\right)_{(p, q)}= \frac{\delta_{p, q}}{(2 \pi)^{2}}\times\\&\hspace{-3mm} \left(\begin{array}{cc}
0 & \hspace{-5mm} - \slashed p^{T}\left[1+r\left(\frac{q^{2}}{k^{2}}\right)\right]+i \slashed d^{T}_k-i {m}_{1,k}{-im_{2,k}\sigma_3}\hspace{-1mm} \\
\hspace{-1mm}- \slashed p\left[1+r\left(\frac{q^{2}}{k^{2}}\right)\right]-i \slashed d_k+i {m}_{1,k}{+im_{2,k} \sigma_3} & 0
\end{array}\right).
\end{aligned}
\end{equation}
If $B$ and $C$ are invertible matrices, we have that

\begin{equation}\label{matrices}
\left[\begin{array}{cc}
0 & B \\
C & 0
\end{array}\right]^{-1}=\left[\begin{array}{cc}
0 & C^{-1} \\
B^{-1} & 0
\end{array}\right],
\end{equation}
which we can apply to the inverted regularized propagator together with inversion of linear combinations of sigma matrices

\begin{equation}\label{PauliInverses}
(z_{0} \sigma_{0}+z_{1} \sigma_{1}+z_{2} \sigma_{2}+z_{3} \sigma_{3})^{-1}=\frac{z_{0} \sigma_{0}-z_{1} \sigma_{1}-z_{2} \sigma_{2}-z_{3} \sigma_{3}}{z_{0}^{2}-z_{1}^{2}-z_{2}^{2}-z_{3}^{2}},
\end{equation}
 provided that $z_{0}^{2}-z_{1}^{2}-z_{2}^{2}-z_{3}^{2} \neq 0$. Using those identities, the field-independent propagator is found to be

\begin{equation}\label{Propagator}
\left(\Gamma_{k, 0}^{(2)}+R_{k}\right)^{-1} =(2 \pi)^{2} \delta_{p, q}\left(\begin{array}{cc}0 & \frac{\slashed\alpha_{k}-i {m}_{1,k}{+im_{2,k} }}{\alpha_{k}^{2}+\mu_k^2} \\ \frac{\slashed{\beta}_{k}^{T}+i {m}_{1,k}{-i m_{2,k}}}{\beta_{k}^{2}+\mu_k^2} & 0\end{array}\right) =(2 \pi)^{2} \hat{G}_{0} \delta_{p, q}, 
\end{equation}
where $\slashed{\alpha}_{k}=- \slashed q\left[1+r\left(\frac{q^{2}}{k^{2}}\right)\right]-i \slashed d_{k} \equiv a_{k} \slashed q-i \slashed d_{k}$ and $\slashed{\beta}_{k}=$ $- \slashed q\left[1+r\left(\frac{q^{2}}{k^{2}}\right)\right]+i \slashed d_{k} \equiv a_{k}\slashed q+i \slashed d_{k} ; \quad \alpha^{2}=\alpha_{1}^{2}+\alpha_{2}^{2} \quad$ and $\beta^{2}=\beta_{1}^{2}+\beta_{2}^{2}$ and $\mu_k^2\equiv m_{1,k}^2-m_{2,k}^2$.\\


 The field-dependent part of the propagator $\Delta\Gamma_k^{(2)} \propto \lambda_k$ remains unchanged with respect to the Hermitian case, namely directly follows the computation in~\cite{Russian_FRGThirring}. We directly obtain
\begin{equation}\label{FieldDependent}
\begin{aligned}
\Delta \Gamma_{k}^{(2)}=& 2(2 \pi)^{2}\lambda_k  \times\left(\begin{array}{cc}
-\bar{\psi}^{T} \bar{\psi} & \bar{\psi}^{T} \psi^{T}+\psi^{T} \bar{\psi}^{T} \\
\psi \bar{\psi}+\bar{\psi} \psi & -\psi \psi^{T}
\end{array}\right) \delta_{p, q} \\
=& 2(2 \pi)^{2}\lambda_k \hat{G}_{1} \delta_{p, q},
\end{aligned}
\end{equation}
where the derivative is evaluated at constant background fields. That means we evaluate $\Delta \Gamma_{k}^{(2)}$  at $\psi_{\mathbf{p}}=(2 \pi)^{2} \psi \delta(\mathbf{p})$ in momentum space and hence on the right-hand side, $\psi(\bar{\psi})$ are constants~\cite{63}.\\


 By combining Eqs.~(\ref{Wetterich}) and~(\ref{InvertedRegProp}), we now expand the flow equation in powers of the fields 
\begin{equation}\label{ExpansionWetterich}
\begin{aligned}
\partial_{k} \Gamma_{k}=&-\frac{1}{2} \operatorname{Tr}\left[\tilde{\partial}_{k} \log \left(\Gamma_{k, 0}^{(2)}+R_{k}\right)\right]-\frac{1}{2} \operatorname{Tr}\left[\tilde{\partial}_{k} \frac{\Delta \Gamma_{k}^{(2)}}{\Gamma_{k, 0}^{(2)}+R_{k}}\right] \\
&+\frac{1}{4} \operatorname{Tr}\left[\tilde{\partial}_{k}\left(\frac{\Delta \Gamma_{k}^{(2)}}{\Gamma_{k, 0}^{(2)}+R_{k}}\right)^{2}\right]+\cdots
\end{aligned}
\end{equation}
The beta functions for the various coupling strengths may be obtained by comparing the coefficients of each fermion monomial of the right-hand side of Eq.~(\ref{ExpansionWetterich}) with the coupling terms included in the anzatz~(\ref{EffectiveAction}). We now compute the beta functions for the two-fermion and four-fermion terms in Eq.~(\ref{ExpansionWetterich}).\\


 We focus here on the RG flow equations for the coupling strengths appearing in the quadratic part of the action.
The only term contributing in the expansion~(\ref{ExpansionWetterich}) is given by
\begin{equation}\label{TwoFermionBeta}
-\frac{1}{2} \operatorname{Tr}\left[\frac{\Delta \Gamma_{k}^{(2)}}{\Gamma_{k, 0}^{(2)}+R_{k}}\right]=-\lambda_k \Omega \int \frac{d^{2} q}{(2 \pi)^{2}} \operatorname{Tr}\left[\hat{G}_{1} \hat{G}_{0}\right],
\end{equation}
where $\Omega$ is the volume of the system and the definition of $\hat{G}_{1}$ is found in Eq.~(\ref{FieldDependent}). Using $\mathrm{Tr}(\psi\bar{\psi}A) = -(\bar{\psi} A\psi)$, valid for Grassmann fields, and some algebra, we compute
\begin{equation}\label{B2}
\begin{aligned}
\operatorname{Tr}\left[\hat{G}_{1} \hat{G}_{0}\right]=&-\frac{\alpha_{k \mu}}{\alpha_{k}^{2}+\mu_{k}^{2}}\left(\bar{\psi} \sigma_{\mu} \psi\right)+\frac{\beta_{k \mu}}{\beta_{k}^{2}+\mu_{k}^{2}}\left(\bar{\psi} \sigma_{\mu} \psi\right)       \\&-i\left(\frac{1}{\alpha_{k}^{2}+\mu_{k}^{2}}+\frac{1}{\beta_{k}^{2}+\mu_k^2}\right)(  {m}_{1,k}(\bar{\psi} \psi)+ { {m}_{2,k}(\bar{\psi} \sigma_3\psi)})             \\
=&-2 i {m}_{1,k} \frac{a_{k}^{2} q^{2}-d_{k}^{2}+\mu_{k}^{2}}{\left(a_{k}^{2} q^{2}-d_{k}^{2}+\mu_k^2\right)^{2}+4 a_{k}^{2} d_{k}^{2} q_{1}^{2}}(\bar{\psi} \psi)\\&{ -2 i {m}_{2,k} \frac{a_{k}^{2} q^{2}-d_{k}^{2}+\mu_k^2}{\left(a_{k}^{2} q^{2}-d_{k}^{2}+\mu_k^2\right)^{2}+4 a_{k}^{2} d_{k}^{2} q_{1}^{2}}(\bar{\psi} \sigma_3\psi)}\\ &+2 i d_{k} \frac{a_{k}^{2} q^{2}-d_{k}^{2}+\mu_k^2-2 a_{k}^{2} q_{1}^{2}}{\left(a_{k}^{2} q^{2}-d_{k}^{2}+\mu_k^2\right)^{2}+4 a_{k}^{2} d_{k}^{2} q_{1}^{2}}\left(\bar{\psi} \sigma_{1} \psi\right) \\
&-4 i d_{k} \frac{a_{k}^{2} q_{1} q_{2}}{\left(a_{k}^{2} q^{2}-d_{k}^{2}+\mu_k^2\right)^{2}+4 a_{k}^{2} d_{k}^{2} q_{1}^{2}}\left(\bar{\psi} \sigma_{2} \psi\right).
\end{aligned}
\end{equation}
After integration over the momentum $\mathbf{q}$ in Eq.~(\ref{TwoFermionBeta}), the fourth term ($\propto q_1q_2$) in Eq.~(\ref{B2}) drops out. This  yields
\begin{equation}\label{B3}
\begin{aligned}
-\frac{1}{2} \operatorname{Tr}\left[\tilde{\partial}_{k} \frac{\Delta \Gamma_{k}^{(2)}}{\Gamma_{k, 0}^{(2)}+R_{k}}\right]&=+2 i \Omega {m}_{1,k}\lambda_k \tilde{\partial}_{k} \mathcal{L}_{1}(\bar{\psi} \psi)+2 i \Omega {m}_{2,k}\lambda_k \tilde{\partial}_{k} \mathcal{L}_{1}(\bar{\psi} \sigma_3\psi)\\ &\quad -2 i \Omega d_{k}\lambda_k \tilde{\partial}_{k} \mathcal{L}_{2}\left(\bar{\psi} \sigma_{1} \psi\right),
\end{aligned}
\end{equation}
where the threshold functions $\mathcal{L}_{1}$ and $\mathcal{L}_{2}$ are given by
\begin{equation}\label{B4B5}
\begin{aligned}
\mathcal{L}_{1} &=\int \frac{d^{2} q}{(2 \pi)^{2}} \frac{a_{k}^{2} q^{2}-d_{k}^{2}+\mu_{k}^{2}}{\left(a_{k}^{2} q^{2}-d_{k}^{2}+\mu_{k}^{2}\right)^{2}+4 a_{k}^{2} d_{k}^{2} q_{1}^{2}}, \\
\mathcal{L}_{2} &=\int \frac{d^{2} q}{(2 \pi)^{2}} \frac{a_{k}^{2} q^{2}-d_{k}^{2}+\mu_{k}^{2}-2 a_{k}^{2} q_{1}^{2}}{\left(a_{k}^{2} q^{2}-d_{k}^{2}+\mu_{k}^{2}\right)^{2}+4 a_{k}^{2} d_{k}^{2} q_{1}^{2}}.
\end{aligned}
\end{equation}
From the ansatz for the quadratic part of the effective action~(\ref{EffectiveAction}), we have that
\begin{equation}\label{B6}
\partial_{k} \Gamma_{k}=-i \Omega\left(\partial_{k} d_{k}\right)\left(\bar{\psi} \sigma_{1} \psi\right)+i \Omega\left(\partial_{k} {m}_{1,k}\right) \bar{\psi} \psi +i \Omega\left(\partial_{k} {m}_{2,k}\right) (\bar{\psi} \sigma_3\psi) .
\end{equation}
Identifying the coefficients of the quadratic contributions in Eqs.\,\eqref{B3} and\,\eqref{B6} to the flow equations yields the expressions for the beta functions of the mass $m_{1,k}$, its $\gamma_5$-dependent counterpart $m_{2,k}$ and the external gauge field $d_k$
\begin{equation}\label{B7B8}
\begin{aligned}
\partial_{k} m_{1,k} &=2\lambda_k m_{1,k} \tilde{\partial}_{k} \mathcal{L}_{1}, \\
{\partial_{k} m_{2,k}} &=2\lambda_k m_{2,k} \tilde{\partial}_{k} \mathcal{L}_{1}, \\
\partial_{k} d_{k} &=2\lambda_k d_{k} \tilde{\partial}_{k} \mathcal{L}_{2}.
\end{aligned}
\end{equation}
Now turning to the four-fermion beta function, we have
\begin{equation}\label{FourFermionBeta}
\frac{1}{4} \operatorname{Tr}\left[\tilde{\partial}_{k}\left(\frac{\Delta \Gamma_{k}^{(2)}}{\Gamma_{k, 0}^{(2)}+R_{k}}\right)^{2}\right]=\lambda_k^{2} \Omega \int \frac{d^{2} q}{(2 \pi)^{2}} \operatorname{Tr}\left[\hat{G}_{1} \hat{G}_{0} \hat{G}_{1} \hat{G}_{0}\right].
\end{equation}
We must evaluate
\begin{equation}\label{C2}
\begin{aligned}
&\operatorname{Tr}\left[\hat{G}_{1} \hat{G}_{0} \hat{G}_{1} \hat{G}_{0}\right]=\\ &\quad \quad \left\{-\operatorname{det}\left(\frac{\slashed\alpha_{k}-i {m}_{1,k}+i m_{2,k}\sigma_3}{\alpha_{k}^{2}+\mu_{k}^{2}}\right)-\operatorname{det}\left(\frac{\slashed\beta_{k}-i {m}_{1,k}+i m_{2,k}\sigma_3}{\beta_{k}^{2}+\mu_{k}^{2}}\right)\right.\\ & \quad\quad\quad\quad\quad\quad\quad\quad\left.
+2 \mathcal{X}\left(\frac{\slashed\alpha_{k}-i {m}_{1,k}+i m_{2,k}\sigma_3}{\alpha_{k}^{2}+\mu_{k}^{2}}, \frac{\slashed\beta_{k}+i {m}_{1,k}+i m_{2,k}\sigma_3}{\beta_{k}^{2}+\mu_{k}^{2}}\right)\right\}(\bar{\psi} \psi)^{2},
\end{aligned}
\end{equation}
where $\mathcal{X}$ takes matrix arguments and is defined through
\begin{equation}\label{C3}
\mathcal{X}(\hat{M}, \hat{N})=\frac{1}{2}\left(M_{11} N_{22}-M_{12} N_{21}-M_{21} N_{12}+M_{22} N_{11}\right).
\end{equation}
Evaluating the quartic part of ansatz~(\ref{EffectiveAction}) for constant fields, we have
\begin{equation}\label{ConstantFieldInteraction}
\Gamma_{k}=\Omega\lambda_k(\bar{\psi} \psi)^{2} .
\end{equation}
The flow for the coupling constant $\lambda_{k}$ is obtained by using Eqs.~(\ref{FourFermionBeta}) and~(\ref{ConstantFieldInteraction}) and comparing both sides of Eq.~(\ref{ExpansionWetterich})
\begin{equation}\label{C5}
\partial_{k} \lambda_{k}=\lambda_{k}^{2} \tilde{\partial}_{k} \mathcal{L}_{3},
\end{equation}
where the threshold function $\mathcal{L}_{3}$ is given by
\begin{equation}\label{L3}
\begin{aligned}
\hspace{-2mm}\mathcal{L}_{3}=\hspace{-1mm}\int \frac{d^{2} q}{(2 \pi)^{2}}\hspace{-1mm}\left\{-\operatorname{det}\left(\frac{\slashed\alpha_{k}-i {m}_{1,k}+i m_{2,k}\sigma_3}{\alpha_{k}^{2}+\mu_{k}^{2}}\right)-\operatorname{det}\left(\frac{\slashed{\beta}_{k}-i {m}_{1,k}+i m_{2,k}\sigma_3}{\beta_{k}^{2}+\mu_{k}^{2}}\right)\right.\\\quad\quad\quad\quad\left.+2 \mathcal{X}\left(\frac{\alpha_{k}-i {m}_{1,k}+i m_{2,k}\sigma_3}{\alpha_{k}^{2}+\mu_{k}^{2}}, \frac{\slashed{p}_{k}+i {m}_{1,k}+i m_{2,k}\sigma_3}{\beta_{k}^{2}+\mu_{k}^{2}}\right)\right\}.
\end{aligned}
\end{equation}
Summarizing, we have so far obtained Eqs.\,\eqref{B7B8} and\,\eqref{C5}, which express the beta functions of the various coupling strengths in terms of the regulator scale dependence of momentum integrals given by the threshold functions $\mathcal{L}_{1,2,3}$.  Practical derivations of the flow equations may be performed using the sharp cutoff regulator
\begin{equation}\label{SharpCutoff}
r\left(\frac{q^{2}}{k^{2}}\right)=\left\{\begin{array}{ll}
\infty, & q^{2}<k^{2} \\
0, & q^{2}>k^{2}
\end{array}\right. ,
\end{equation}
which renders most easy the explicit analytical evaluation of the threshold functions $\mathcal{L}_{1,2,3}$. Detailed calculations are given in section~\ref{Threshold} and yield
\begin{equation}
\begin{aligned}
k \tilde{\partial}_{k} \mathcal{L}_{1}&=-\frac{\left(k^{2}-d_{k}^{2}+\mu_{k}^{2}\right)}{4 \pi d_{k}^{2}} \frac{1}{\sqrt{\left[2\left(\frac{k^{2}-d_{k}^{2}+\mu_{k}^{2}}{2 k d_{k}}\right)^{2}+1\right]^{2}-1}}\\
k \tilde{\partial}_{k} \mathcal{L}_{2}&=\left[1+\frac{1}{2 d^{2}}\left(k^{2}-d_k^{2}+\mu_{k}^{2}\right)\right] k \tilde{\partial}_{k} \mathcal{L}_{1}+\frac{k^{2}}{4 \pi d^{2}_k}\\
k \tilde{\partial}_{k} \mathcal{L}_{3}&=\frac{4\left(\mu_{k}^{2}-d_{k}^{2}\right)}{\left(k^{2}-d_{k}^{2}+\mu_{k}^{2}\right)} k \tilde{\partial}_{k} \mathcal{L}_{1}.
\end{aligned}
\end{equation}

We now express the latter in terms of the dimensionless coupling strengths $\bar{\lambda}=\lambda_{k} / 2, \bar{m}_1=(2 k)^{-1} {m}_{1,k}$, $\bar{m}_2=(2 k)^{-1} {m}_{2,k}$, $\bar{\mu}^2=\bar{m}_1^2-\bar{m}_2^2$, $\bar{d}=(2 k)^{-1} d_{k}$,  defined in units of the running scale $k$. The resulting RG equations read
\begin{equation}\label{FlowFinal}
\begin{aligned}
\partial_\tau \bar{m}_1&=\bar{m}_1+\frac{\bar{m}_1 \bar{\lambda}}{4 \pi \bar{d}^{2}} \frac{\left(1-4 \bar{d}^{2}+4 \bar{\mu}^2\right)}{\sqrt{\left[2\left(\frac{1-4 \bar{d}^{2}+4\bar{\mu}^2}{4 \bar{d}}\right)^{2}+1\right]^{2}-1}}\\
\partial_\tau \bar{m}_2&=\bar{m}_2+\frac{\bar{m}_2 \bar{\lambda}}{4 \pi \bar{d}^{2}} \frac{\left(1-4 \bar{d}^{2}+4 \bar{\mu}^2\right)}{\sqrt{\left[2\left(\frac{1-4 \bar{d}^{2}+4\bar{\mu}^2}{4 \bar{d}}\right)^{2}+1\right]^{2}-1}}\\
\partial_\tau \bar{d}&=\bar{d}-\frac{\bar{\lambda}}{4 \pi \bar{d}}+\frac{\bar{\lambda}}{4 \pi \bar{d}} \frac{\left(1-4 \bar{d}^{2}+4 \bar{\mu}^2\right)\left(1+\frac{1}{8 \bar{d}^{2}}\left[1-4 \bar{d}^{2}+4 \bar{\mu}^2\right]\right)}{\sqrt{\left[2\left(\frac{1-4 \bar{d}^{2}+4 \bar{\mu}^2}{4 \bar{d}}\right)^{2}+1\right]^{2}-1}} \\
\partial_\tau \bar{\lambda}&=-\frac{2 \bar{\lambda}^{2}}{\pi}\left(1-\frac{\bar{\mu}^2}{\bar{d}^{2}}\right) \frac{1}{\sqrt{\left[2\left(\frac{1-4 \bar{d}^{2}+4 \bar{\mu}^2}{4 \bar{d}}\right)^{2}+1\right]^{2}-1}},
\end{aligned}
\end{equation}
where the "RG time" $\tau$ is defined as $|\tau|=\ln (\Lambda / k)$. As explained in the text,  the renormalization of $\bar{m}_1$ and $\bar{m}_2$ are not independent, in particular we have
 
  \begin{equation}
        \frac{(\partial_\tau \bar{m}_1)}{\bar{m}_1} = \frac{(\partial_\tau \bar{m}_2)}{\bar{m}_2}, 
  \end{equation}
  and hence the ratio between $\bar{m}_1$ and $\bar{m}_2$ is fixed under renormalization
 \begin{equation}
        \partial_\tau\left(\frac{ \bar{m}_1}{\bar{m}_2}\right) = \frac{(\partial_\tau \bar{m}_1)}{\bar{m}_2}-\bar{m}_1\frac{(\partial_\tau \bar{m}_2)}{\bar{m}_2^2} = \left(\frac{ \bar{m}_1}{\bar{m}_2}\right)\left(\frac{(\partial_\tau \bar{m}_1)}{\bar{m}_1} - \frac{(\partial_\tau \bar{m}_2)}{\bar{m}_2}\right) =0, 
  \end{equation}
as reported in Eq.~(\ref{Ratio}). Equations~(\ref{FlowFinal}) can be reexpressed entirely in terms of the propagating mass $\mu$: using
 
 \begin{equation}
 \begin{aligned}
     \tfrac{1}{2}\partial_\tau \bar{\mu}^2  &= \tfrac{1}{2}\partial_\tau (\bar{m}_1^2-\bar{m}_2^2) =  \bar{m}_1 \partial_\tau (\bar{m}_1) -  \bar{m}_2 (\partial_\tau \bar{m}_2)\\
     \tfrac{1}{2}\partial_\tau \bar{d}^2 &= \bar{d}\partial_\tau \bar{d},
 \end{aligned}
 \end{equation}
and plugging in Eqs.~(\ref{FlowFinal}) leads to the flow equations~(\ref{FlowGauge}) upon which we base our analysis.

\section{Threshold functions} \label{Threshold}
The threshold functions defined in Eqs.~(\ref{B4B5}) and~(\ref{L3}) contain the details of the scale dependence of the regulator in the regularized propagator. In the flow equations, $\tilde{\partial}_{k}$ is defined to act on the regulator's $k$-dependence. The sharp cut-off regulator~(\ref{SharpCutoff}) remarkably enables the explicit analytical evaluation of all threshold integrals. In polar coordinates, we find
\begin{equation}
\tilde{\partial}_{k} \mathcal{L}_{1}=\tilde{\partial}_{k} \int_{k}^{\Lambda} \frac{d q}{(2 \pi)^{2}} \int_{0}^{2 \pi} d \phi \frac{q\left(q^{2}-d_{k}^{2}+\mu_k^{2}\right)}{\left(q^{2}-d_{k}^{2}+\mu_k^{2}\right)^{2}+4 d_{k}^{2} q^{2} \cos ^{2} \phi},
\end{equation}
where $\Lambda$ is the UV cutoff, and $a=-1$ for the cutoff~(\ref{SharpCutoff}).
The $k$-dependence is only present for values of momenta $q^2$ smaller than $k^2$ in the momentum integral. Using

\begin{equation}
\tilde{\partial}_{k} \mathcal{L}_{1}=\tilde{\partial}_{k} \int_{k}^{\Lambda} dq\,f(q) = \tilde{\partial}_{k} \int_{0}^{\Lambda} dq\,f(q)\theta(k-q) = \int_{0}^{\Lambda} dq\,f(q)(-\delta(k-q)) = -f(k) ,
\end{equation}
one obtains
\begin{equation}
k \tilde{\partial}_{k} \mathcal{L}_{1}=-\frac{1}{8 \pi^{2}} \frac{\left(k^{2}-d_{k}^{2}+\mu_k^{2}\right)}{d^{2}} \int_{0}^{2 \pi} \frac{d \phi}{2\left(\frac{k^{2}-d_{k}^{2}+\mu_k^{2}}{2 k d_{k}}\right)^{2}+1+\cos 2 \phi}.
\end{equation}
Performing the integral yields the RG running of $\mathcal{L}_{1}$
\begin{equation}
k \tilde{\partial}_{k} \mathcal{L}_{1}=-\frac{\left(k^{2}-d_{k}^{2}+\mu_{k}^{2}\right)}{4 \pi d_{k}^{2}} \frac{1}{\sqrt{\left[2\left(\frac{k^{2}-d_{k}^{2}+\mu_{k}^{2}}{2 k d_{k}}\right)^{2}+1\right]^{2}-1}}.
\end{equation}
Similarly, the scale dependence of $\mathcal{L}_{2}$ is given by
\begin{equation}
k \tilde{\partial}_{k} \mathcal{L}_{2}=\left[1+\frac{1}{2 d^{2}}\left(k^{2}-d_k^{2}+\mu^{2}_k\right)\right] k \tilde{\partial}_{k} \mathcal{L}_{1}+\frac{k^{2}}{4 \pi d^{2}_k}.
\end{equation}
To find the scale derivative of $\mathcal{L}_{3}$, we make use of the following identities
\begin{equation}
\begin{aligned}
\operatorname{det}\left(\frac{\slashed\alpha_{k}-i m_{1,k}+i {m}_{2,k}\sigma_3}{\alpha_{k}^{2}+\mu_{k}^{2}}\right)&=-\frac{1}{\alpha_{k}^{2}+\mu_{k}^{2}}\\ \operatorname{det}\left(\frac{\slashed\beta_{k}+i {m}_{1,k}-i {m}_{2,k}\sigma_3}{\beta_{k}^{2}+\mu_{k}^{2}}\right)&=-\frac{1}{\beta_{k}^{2}+\mu_{k}^{2}},
\end{aligned}
\end{equation}
and
\begin{equation}
\mathcal{X}\left(\frac{\slashed\alpha_{k}-i m_{1,k}+i m_{2,k}\sigma_3}{\alpha_{k}^{2}+\mu_{k}^{2}}, \frac{\slashed\beta_{k}+i {m}_{1,k}-i {m}_{2,k}\sigma_3}{\beta_{k}^{2}+\mu_{k}^{2}}\right)=\frac{\mu_{k}^{2}-\alpha_{1 k} \beta_{1 k}-\alpha_{2 k} \beta_{2 k}}{\left(\alpha_{k}^{2}+\mu_k^2\right)\left(\beta_{k}^{2}+\mu_k^2\right)}.
\end{equation}
Therefore, 
\begin{equation}
\begin{aligned}
&-\operatorname{det}\left(\frac{\slashed\alpha_{k}-i {m}_{1,k}+i {m}_{2,k}\sigma_3}{\alpha_{k}^{2}+\mu_{k}^{2}}\right)-\operatorname{det}\left(\frac{\slashed{\beta}_{k}-i {m}_{1,k}+i {m}_{2,k}\sigma_3}{\beta_{k}^{2}+\mu_{k}^{2}}\right)\\& \quad\quad  +2 \mathcal{X}\left(\frac{\slashed{\alpha}_{k}-i {m}_{1,k}+i {m}_{2,k}\sigma_3}{\alpha_{k}^{2}+\mu_{k}^{2}},        \frac{\slashed{\beta}_{k}+i {m}_{1,k}-i {m}_{2,k}\sigma_3}{\beta_{k}^{2}+\mu_{k}^{2}}\right) \\
& \quad\quad\quad\quad  =\frac{4 \mu_{k}^{2}+\left(\alpha_{1 k}-\beta_{1 k}\right)^{2}+\left(\alpha_{2 k}-\beta_{2 k}\right)^{2}}{\left(\alpha_{k}^{2}+\mu_{k}^{2}\right)\left(\beta_{k}^{2}+\mu_{k}^{2}\right)} .
\end{aligned}
\end{equation}
The choice of the sharp cutoff~(\ref{Regulator}), for which $\alpha_{1 k}-\beta_{1 k}=-2 i d_{k}$ and $\alpha_{2 k}-\beta_{2 k}=0$, yields
\begin{equation}
\frac{4\left(\mu_{k}^{2}-d_{k}^{2}\right)}{\left(q^{2}-d_{k}^{2}+\mu_{k}^{2}\right)^{2}+4 q_{1}^{2} d_{k}^{2}} .    
\end{equation}
Inserting this result in Eq.~(\ref{L3}), we obtain
\begin{equation}
\mathcal{L}_{3}=\frac{1}{\pi^{2}} \int_{k}^{\Lambda} d q q \int_{0}^{2 \pi} d \phi \frac{\left(\mu_{k}^{2}-d_{k}^{2}\right)}{\left(q^{2}-d_{k}^{2}+\mu_{k}^{2}\right)^{2}+4 q^{2} d_{k}^{2} \cos ^{2} \phi}.
\end{equation}
In particular, one finds that
\begin{equation}
k \tilde{\partial}_{k} \mathcal{L}_{3}=\frac{4\left(\mu_{k}^{2}-d_{k}^{2}\right)}{\left(k^{2}-d_{k}^{2}+\mu_{k}^{2}\right)} k \tilde{\partial}_{k} \mathcal{L}_{1}.
\end{equation}

\twocolumngrid


%


\end{document}